\documentclass[twocolumn]{article}

\usepackage{graphicx}

\newcommand{\be}{\begin{equation}}
\newcommand{\ee}{\end{equation}}
\renewcommand{\k}[2]{\frac{#1}{#2}}

\newcommand{\pab}[2]{\frac{\p #1}{\p #2}}

\def\p{\partial}
\def\s{\,\,\,\,}

\def\={\approx}

\def\v{\lambda}
\def\ra{\rightarrow}
\def\r{{\cal R}}
\def\ga{\Xi}

\title{\large \bf A Unified Approach to Mechanical Compaction, Pressure Solution,
Mineral Reactions  and the Temperature Distribution in Hydrocarbon Basins }

\author{Xin-She Yang \\
  Department of  Applied Mathematics and  Department of Fuel and Energy \\
  University of Leeds, LEEDS LS2 9JT, UK 
}

\date{}
\begin{document}

\maketitle

\begin{abstract}

In modelling sediment compaction and mineral reactions, the rheological
behaviour of sediments is typically considered as poroelastic or
purely viscous. In fact, compaction due to pressure solution
and mechanical processes in porous media is far more complicated.
A generalised  model of viscoelastic compaction and the smectite to illite
mineral reaction in hydrocarbon basins is presented.
A one-step dehydration model of the
mineral reaction is assumed.  The obtained nonlinear governing equations
are solved numerically and different combinations of physical
parameters are used to simulate realistic situations in typical
sedimentary basins.    Comparison of numerical simulations
with real data has shown very good agreement with respect to both the
porosity profile and the mineral reaction. \\


\end{abstract}

\noindent Published in {\it  Tectonophysics}, {\bf 330}, 141-151 (2001).

\newpage

\section{Introduction}

When a well-bore is being drilled for oil exploration, drilling mud
is used in the hole to maintain its integrity and safety.
The mud density must be sufficient to prevent collapse of the hole, but not
so high that hydrofracturing of the surrounding rock occurs. Both these
effects depend on the pore fluid pressure in the rock, and drilling
problems occur in regions where abnormal pore pressure occurs.
Such overpressuring can substantially affect oil-drilling
rates and even cause serious blowouts during drilling.  Therefore, an
industrially important objective is to predict overpressuring
before drilling and to identify its precursors during drilling.
Another related objective is to predict reservoir quality and
hydrocarbon migration. An essential step to achieve such objectives
is the scientific understanding of overpressuring mechanisms and the
evolutionary history of post-depositional sediments such as  shales.

The extent of compaction is strongly influenced by burial history and
the lithology of sediments. The freshly deposited loosely packed
sediments tend to evolve, like an open system, towards a closely packed
grain framework during the initial stages of burial compaction and this is
accomplished by the processes of grain slippage, rotation, bending and
brittle fracturing. Such reorientation processes are collectively
referred to as {\em mechanical compaction},
and generally take place in  the first 1 - 2 km of burial.
After this initial porosity loss,
further porosity reduction is accomplished by processes of
{\em chemical compaction} such as pressure solution  (Rutter, 1983;
Tada and Siever, 1989). Overpressuring may develop from a
{\em non-equilibrium compaction} environment (Hedberg, 1936;
Gibson et al, 1967; Audet and Fowler, 1992; Connolly, 1997;
Connolly and Podladchikov, 1998; Fowler and Yang, 1998).
A rapidly accumulating basin is unable to expel pore fluids sufficiently
rapidly due to the weight of overburden rock. The development of
overpressuring retards compaction, resulting in higher porosity,
permeability and low thermal conductivity than are normal for a
given depth. These changes alter the structural and stratigraphic features
of sedimentary units and provide the potential for hydrocarbon migration.
Rieke and Chilingarian (1974) gave a detailed review on this topic.
Audet and Fowler (1992) also reviewed compaction briefly  and
Connolly and Podladchikov (2000) provided a more recent comprehensive review.

Pressure solution has been considered as an important
process in deformation and porosity change during compaction
in sedimentary rocks (Bathurst, 1971;Tada and Siever, 1989).
Pressure solution refers to a process by which grains dissolve at
intergranular contacts under non-hydrostatic stress
and reprecipitate in pore spaces, thus resulting in compaction.
The solubility of minerals increases with increasing effective stress at
grain contacts. Pressure dissolution at grain contacts is therefore a
compactional response of the sediment during burial in an attempt to
increase the grain contact area so as to distribute the effective
stress over a larger surface. There has been renewed
interest on pressure solution mechanisms because of its important role
in the diagenesis of sedimentary rocks and its relation to rock deformation.
Two common types of pressure solutions occur in  nature.
The intergranular pressure solution was first studied by Sorby (1863) based
while the stylolitization was first described by Fuchs (1894) followed
by many authors such as Stockdale (1922), Dunnington (1954)
and Heald (1955). Tada and Siever (1989) gave a comprehensive review
on pressure solution and some recent literature can be found in
recent papers (Birchwood  and Turcotte, 1994; Schneider, 1996;
Connolly and Podladchikov, 1998; Fowler and Yang, 1999; Yang, 2000a).
Pressure solution process is typically assumed to viscous
(Weyl, 1959; Rutter, 1983; Birchwood and Turcotte, 1994;
Fowler and Yang, 1999)  and it is usually
referred to as viscous compaction, viscous creep or pressure
solution creep. Its rheological constitutive relation
(or compaction relation) is often written as a relationship
between effective stress and strain rate. However, the more natural
and realistic rheology shall be viscoelastic (Connolly and Podladchikov, 1998;
Revil, 1999; Yang, 2000c) and this is the approach we will use in the
present paper.

Mineral reactions are observed worldwide in sedimentary hydrocarbon
basins. The close spatial correlations between smectite
disappearance and illite formation imply   the existence  of
the smectite-illite reaction.
This transformation is one of the most important in clastic rocks.
The reaction has received much attention but the nature of both the
illite/smectite mixed-layer and the reaction mechanism are still
under discussion (Eberl and Hower 1976;
Velde and Vasseur, 1992; Abercrombie et al., 1994).
Measured rates of this process
in the laboratory (Eberl and Hower 1976)  suggest that at elevated
temperature, the transformation
proceeds very fast from the geological
point of view. On the other hand, observations suggest that
this reaction is initiated relatively suddenly at a temperature
of $90^{0} C$, but then takes place
gradually over a depth of several hundred meters, which suggests a
time scale of the order of a million years.  The transformation
can be considered a simple dehydration reaction. In fact,
the mechanism of such a mineral reaction is rather more complicated
and is not completely understood. The mineral reaction
may consist of dissolution of montmorillonite in free
pore water and subsequent precipitation of silica as illite.
The following dissolution-precipitation reactions
have been proposed by Yang (2000b) and Fowler (2000)\\
{\em Smectite dissolution}
\begin{equation}
M^{s} (\mbox{smectite})  \rightarrow  [-Si^{l}-]+n [H_{2}O],
\end{equation}

\noindent {\em K-feldspar dissolution}
\begin{equation}
[\mbox{K-feldspar}] \rightarrow [K^{+l}]+[AlO_{2}^{-l}-]+ [Si
O_{2}^{l}],
\end{equation}

\noindent {\em Illite precipitation}
\begin{equation}
[K^{+l}] + [AlO_{2}^{-l}-]+  [-Si^{l}-] \rightarrow
I^{s} (\mbox{illite})+[SiO_{2}^{l}],
\end{equation}

\noindent {\em Quartz dissolution and precipitation}
\begin{equation}
[Si O_{2}^{l}] \leftrightarrow [\mbox{quartz}],
\end{equation}
where $s, l$ denote solid and liquid phase. $[-Si-]$ is an aqueous
silica combination in such forms as $[-(Si_{4})O_{10}(OH)_{2}]$.
$[AlO_{2}^{-L}-]$ is only a  general notation of the combination such as
$[Al(OH)_{4}^{-}]$. The reaction rates in the above dissolution and
reprecipitation mechanism may be quite different at different steps because
dissolution    is    usually enhanced by pressure solution due to the
increased pressure at grain contacts, while
precipipation at free pore space is less enhanced by pressure
(Rimstidt and Barnes, 1980).
In the limiting case, the above four step reactions
can be approximately considered as a one-step (first order)
dehydration process. The first-order dehydration model
is a good approximation in the sense
of describing the extent of progress of the overall smectite-to-illite
transformation without much concern for its detailed geochemical features.
Therefore, we represent it schematically as
\begin{equation}
{\rm smectite} \cdot [H_{2}O] {\stackrel{k_{r}}{\rightarrow}}
{\rm illite}+n_{H_2 O} [H_{2}O] ({\rm free\, water}),
\end{equation}
which was suggested in the earlier works (Eberl and Hower, 1976;
Velde and Vasseur, 1992; Abercromie et al., 1994). This one-step
model was studied briefly by Yang (2000b) and Fowler (2000).

Despite the importance of compaction and mineral reactions for geological
problems, the literature dealing with quantitative modelling is not
extensive. Although qualitative features have been
received much attention in the literature, the
mechanismd are still poorly understood.
A full understanding of the mechanism requires
an interdisciplinary study involving soil mechanics, geochemistry,
geophysics and geology (Hedberg, 1936;
Rieke and Chilingarian, 1974; Tada and Siever, 1989;
Fowler and Yang, 1998; Connolly and Podladchikov,2000).
This paper provides a unified  compaction relation in a form of
a visco-poroelastic relation of Maxwell type. The nonlinear partial
differential equations are then solved numerically and compared with real
borehole log data.

\section{MATHEMATICAL MODEL}

For the convenience of investigating the effect of sediment compaction
and mineral reactions,  the sediment is considered as as a
compressible porous matrix.
Combining  mass conservation of the pore fluid
with Darcy's law leads to model equations of the general
type. Consider a  matrix consisting of four
interdispersed media: free pore water and
three solid phase clay minerals, namely, quartz,
water-rich montmorillonite and dehydrated illite.
Let the volume fractions of the respective media (montmorillonite,
illite,quartz, water) be $\phi_{m}, \phi_{i}, \phi_{c}, \phi$,
so that $\phi_{m}+\phi_{i} +\phi_{c}
+\phi=1$. We assume that all the solids move with  the same averaged
velocity $u^{s}$, while the pore water has velocity $u^{l}$.
The rate at which montmorillonite is transformed is denoted by $r_m$,
the rate of illite formation is $r_i$, and the rate at which water released
from the mineral reaction is $r_w=r_m-r_i$ due to mass conservation.

\noindent {Conservation of mass}
\begin{equation}
\frac{\partial \phi_m }{\partial t}+{\nabla} \cdot
[\phi_m {\bf u}^{s}]=-r_m,  \label{VC:MASS-1}
\end{equation}
\begin{equation}
\frac{\partial \phi_i}{\partial t}+{\nabla} \cdot
[\phi_i {\bf u}^{s}]=r_i,  \label{VC:MASS-2}
\end{equation}
\begin{equation}
\frac{\partial \phi}{\partial t}+{\nabla} \cdot (\phi
{\bf u}^{l})=r_w,
\end{equation}
\begin{equation}
\phi +\phi_c+\phi_i + \phi_m=1,
\end{equation}
{Conservation of Energy}
\[
\frac{\partial }{\partial t}[\rho_l c_l \phi T +\rho_s c_s (1-\phi) T] \] \begin{equation}+\nabla
\cdot \{[\rho_s c_s (1-\phi) {\bf u}^{s} +\rho_{l} c_{l} \phi {\bf u}^{l}] T \} =\nabla \cdot (K_{th} \nabla T),
\end{equation}
{Darcy's law}
\begin{equation}
\phi ({\bf u}^{l}-{\bf u}^{s})=-\frac{k}{\mu}(\nabla p^{l}+
\rho_{l} g {\bf j}),
\end{equation}
 {Force balance}
\begin{equation}
\nabla \cdot \mbox{\boldmath $\sigma$}^{e}-\nabla [{p}^{l}]-\rho
g {\bf j}={\bf 0}, \s \rho=(1-\phi)\rho_s+\phi \rho_l,
\end{equation}
where the momentum equation has been written in the similar form as
Fowler (1990) and Audet and Fowler (1992). This present form
is also equivalent to that given by (McKenzie, 1984) except for
the notation difference.
${\bf u}^{l} $ and ${\bf u}^{s}$ are the
velocities of fluid and solid matrix, $k$ and $\mu$ are the matrix
permeability and the liquid
viscosity,   $\rho_{l} $ and $\rho_{s}$ are the
densities of fluid and solid matrix,   $\mbox{\boldmath $\sigma$}^{e}$ is
the effective stress,
${\bf j}$ is the unit vector pointing vertically upwards.
$p^{l}$ is the pore pressure, and $g$ is the gravitational acceleration.
$K_{th}$ is thermal conductivity, and $T$ is temperature.
$c_s$ and $c_l$ are specific heat of the solid and liquid, respectively.
In formulating the mathematical model, the heat release from
the mineral reaction has been neglected since it is usually very small.
In addition, a rheological constitutive relation, which is often refered as
compaction law, is needed to complete this model and this is described
in the following section \ref{se-21} in more detail.

Combing equation force balance and Darcy's law, we have
\begin{equation}
\phi ({\bf u}^{l}-{\bf u}^{s})
=\frac{k(\phi)}{\mu}[\nabla . \mbox{\boldmath $\sigma$}^{e}-(\rho_s-\rho_l)(1-\phi) g {\bf j}],
\end{equation}
which is a reduced form of Darcy's law and ${\bf j}$ is a unit vector pointing
upward.

\subsection{Poroelasticity, Pressure Solution and Viscoelastic
Compaction} \label{se-21}

Nonlinear compaction models have been formulated in two ways.
The early and most common models assumed an elastic or
poroelastic rheology, and the compaction relation is of the
Athy's type $p_e=p_e
(\phi)$    (Gibson, England \& Hussey, 1967; Smith, 1971;
Sharp, 1976; Wangen, 1992; Audet and Fowler, 1992; Fowler and Yang, 1998).
More recent models assumes a viscous rheology with
a compaction relation of the form
$p_e=-\xi(\phi) \nabla . {\bf u}^s$ where $\xi$ is the bulk viscosity
(Augevine and Turcotte,
1983; Birchwood and Turcotte, 1994;  Fowler and Yang, 1999).
The poroelastic models are valid for the mechanical compaction while
the viscous models are used to simulate the chemical compaction such as
pressure solution. We can thus
generalise the above relations to a viscoelastic compaction law of the
1-D Maxwell type
\be
\pab{u^s}{z}=-\frac{1}{K(\phi)}\frac{D p_e}{Dt}-\frac{1}{\xi(\phi)}p_e,\s
\k{D}{Dt}=\pab{}{t}+u^s \pab{}{z}. \label{equ-777}
\ee
and
\be
K(\phi)=\k{C_s (1-\phi)^2}{\phi}, \s \xi=\xi_0 (\k{\phi_0}{\phi})^n,
\ee
where $C_s$ is a constant or compaction index describing the degree of
compaction of sediments.  $\xi_0$ is the value of viscosity at $\phi=\phi_0$
(the initial porosity).  Equivalently, we would anticipate a
viscoelastic rheology for the medium, involving material
derivatives of tensors, and some care is needed to ensure
that the resulting model is frame indifferent.
That is to say, the rheological relation of
stress-strain should be invariant under the coordinate transformation.
This is not always guaranteed due to the complexity of the rheological
relations (Bird, Armstrong \& Hassager 1977). Fortunately, for
one-dimensional flow, which is always {\em irrotational},
the equation is invariant and all the different equations in corotional and
codeformational frames degenerate into the same form (Yang, 2000c).
In the one-dimensional case we will discuss below, we can take this for
granted.

\section{Non-dimensionalization}

For a 1-D basin, the physical domain is from the basin basement ($z=0$)
to the basin top (ocean floor) and the basin thickness varies with time $t$.
The length-scale $d$ is a typical length which will be defined
\begin{equation}
p=\k{G p_{e}}{(\rho_{s}-\rho_{l}) g d},
\end{equation}
where $G=1+\k{4 \mu}{3 \xi_0}$ is a constant describing sediment
properties and  $p_{e}$ is the effective pressure. The choice of scaling $d$
is in such a way that the dimensionless pressure $p=O(1)$.
Assigning the typical thermal  gradient to be $\gamma$,
we also require that $\Theta=(T-T_0)/(\gamma d)=O(1)$.
Meanwhile,
we  scale $z$ with $d$, $u^{s}$ with
${\dot m}_{s}$, time $t$ with $ d/{\dot m}_{s}$, permeability $k$ with
$k_{0}$, and heat conductivity $K_{th}$ with $K_{0}$ where $\dot m_s, k_0$
and $K_0$ are set to typically observed values.
By writing $k(\phi)=k_{0} k^*$, $z=d z^*$, $K_{th}=K_0 \hat K$, ...,
and dropping the asterisks, we thus have
\begin{equation}
\pab{\phi_m}{t} +\pab{}{z}(\phi_m u^{s})= - \r e^{\beta \Theta} \phi_{m}, ,
\label{VC:MASS-1}
\end{equation}
\begin{equation}
\pab{\phi_i}{t}+ \pab{(\phi_i u^{s})}{z}= \r (1-a) e^{\beta \Theta} \phi_{m},
\label{PHI:U-1}
\end{equation}
\begin{equation}
\pab{\phi}{t}+ \pab{(\phi u^{l})}{z}= a \r \delta e^{\beta  \Theta} \phi_{m},
\label{PHI:U-2}
\end{equation}
\be
\phi+\phi_i+\phi_c+\phi_m=1;
\ee
\[
\frac{\partial}{\partial t}\{[\alpha(1-\phi)+\phi]\Theta\}
+\frac{\partial }{\partial
z}\{[\alpha(1-\phi)u^{s}+\phi u^{l}] \Theta
\} \] \be =\Lambda\frac{\partial}{\partial z}(\hat K \frac{\partial \Theta}{\partial
z}),
\end{equation}
\begin{equation}
\phi (u^{l}-u^{s})=\v k(\phi) [-\pab{p}{z}-(1-\phi) ],
\end{equation}
\be
\pab{u^s}{z}=-\k{C_s \phi}{(1-\phi)^2}\k{ D p}{Dt}-\k{(\phi/\phi_0)^n}{\ga} p,
\ee
where
\[
\lambda=\frac{k_{0} (\rho_{s}-\rho_{l}) g} {\mu {\dot m}_{s}}, \] \be
\Lambda=\frac{K_{0}} {\rho_{l}c_{l} {\dot m}_{s} d},
\s
\ga=\k{\xi_0 \dot m_s G}{(\rho_s-\rho_l) g d^2}.
\ee
\[
{\cal R}=\frac{k^0_{r}d}{\dot m_{s}},\,\,\, a=\frac{n_{w}
M_{w}}{M_{m}}, \,\,\, \beta=\frac{E_{a}}{R T^{2}_{0}}, \] \be
\delta=\k{\rho_s}{\rho_l}, \s \alpha=\frac{\rho_{s} c_{s}}{\rho_{l}c_{l}}.
\end{equation}
and $k_r^0$ and $k_0$ are the reaction rate and permeability at
the basin top, respectively. $C_s$ is a constant coefficient or
compaction index in the known function $K(\phi)$ in equation (\ref{equ-777}).
$n_{w}$ is the molar water released from
one molar smectite; $M_w$ and $M_m$ are the molar weights of water
and smectite. $d$ is the typical length scale in the basin;  $E_a$
is the activation energy of the mineral reaction; $T_0$ is the
temperature at the ocean floor and $R$ is the gas constant.
In addition, $(\phi_0/\phi)^n$ expresses the
dependence of the bulk viscosity $\xi$ on
porosity, which has been
determined empirically to be $n=1.3$ for uncemented sand-like granular
media and $n=2$ for a wide range of cemented  rocks (Paterson, 1995).

The nondimensional parameter $\lambda$ is essentially the ratio of
hydroconductivity to the sedimentation rate and thus characterises the
relative time scales of fluid escape and sedimentation so that $\lambda \ll 1$
is the slow compaction limit while $\lambda \gg 1$ is the fast compaction
limit (Fowler and ang, 1998). Similarly, $\Lambda$ is the ratio of the
thermal conductivity to sedimention rate, and thus controls the progress of
the temperature evolution. The large $\Lambda$ case ($\Lambda \gg 1$)
so that a nearly equilibrium state is reached.
The parameter ${\cal R}$ is the relative
time scales of mineral reaction to sedimentation, and $\beta$ is the reduced
activation energy.  However,
the factor ${\cal R} \exp(\beta \Theta)$ always appears as a
combination in the above model equations, and can be rewritten as
\begin{equation}
{\cal R} \exp(\beta \Theta)= {\rm exp} [ \beta (\Theta - \Theta_{c}) ]
\,\,\,\,{\rm and} \,\,\,\, \Theta_{c}=\frac{1}{\beta} {\rm ln}
\frac{1}{{\cal R}},
\end{equation}
where the new parameter $\Theta_{c}$, which replaces ${\cal
R}$, is a dimensionless critical temperature (with reference to the
surface temperature). In the following discussions, we will see that
the smectite to illite
reaction virtually  takes place in a region called the {\em
reaction window}, at a depth  of $\sim \Theta_{c}$, with its
thickness controlled by $\beta$.

In the above derivation, we have used the requirement of
degenerating  to the poroelastic case  $p= \ln (\phi_0/\phi)
-(\phi_0-\phi)$ when $\ga \ra \infty$.
The constitutive relation for permeability $k(\phi)$ is nonlinear, and
its typical form (Smith, 1971)  is
\be
k(\phi)=(\k{\phi}{\phi_0})^m,
\ee
where $m=8$ is a typical value for shales (Smith, 1971;
Audet and Fowler, 1992). However, the present model can use any value
in the range from $m=3$ to $m=11$ in the numerical simulations.
Initially, the basin thickness is zero or $h(t=0)=0$. During the whole
evolutionary process, new sediment deposits at a constant rate $\dot m_s$
with a constant initial porosity $\phi_0$. The basement rock is assumed
to be impeameable or $u^l=u^s=0$ at $z=0$ while a constant heat flux (-$\hat K
\p \Theta/\p z$) is
introduced at the basement rock. The top boundary is at
constant pressure which can be taken to be zero. Thus,
the boundary conditions     become
\be
u^l=0, \s u^s=0, \s
{\hat K} \pab{\Theta}{z}=-1,
\s  {\rm at } \s z=0,
\label{bbb-1}
\ee
\[ \phi=\phi_{0}, \,\,\,\phi_{m}=\phi_{m0},\,\,\,\, \phi_c=\phi_{c0},\s p=0,\]
\be
\dot h=\dot m(t) + u^s \s {\rm at} \s z=h(t),
\label{bbb-2}
\end{equation}
where $\dot m(t)$ is the nondimensional sedimentation rate.
As the thermal conductivity $\hat K(\phi)$
varies with $\phi$ (Nield,1991), a constant
heat flux will leads to a varying temperature gradient. However,
we will assume $\hat K=const$ (in fitting the borehole data)
because of its weak dependence on porosity and gives a good approximation
in reality (Schneider, 1996).

It is useful  to estimate these parameters  by using values taken
from observations. By using the typical values
of $\rho_{l} \sim 10^{3} \, {\rm kg\, m}^{-3},\,
\rho_{s} \sim 2.5 \times 10^{3} \, {\rm kg\, m}^{-3},\,$
$ k_{0} \sim 10^{-15} -\!\!- 10^{-20}\, {\rm m}^{2}, \, \mu \sim 10^{-3}\,
{\rm N\,s\,
m}^{2}, \, \xi  \sim 1 \times 10^{21}$ N s $ {\rm m}^{-2}, $
$\dot m_{s}  \sim 300\, {\rm m\,\, Ma}^{-1}=1 \times 10^{-11}\,
{\rm m\,\, s}^{-1},\, g \sim 10 {\rm m \,s}^{-2}$, $c_{s} \sim 500 \,{\rm
J}\,\, {\rm Kg}^{-1}\,\, {\rm K}^{-1}$, $c_{l}\sim 4200\,
{\rm J}\, {\rm Kg}^{-1}
\,\,{\rm K}^{-1}$, $K_{0} \sim 1.5 \,\,
{\rm W}\, m^{-1}\,\, K^{-1}$, $T_{0} \sim 280\,\, {\rm K}$,
$ G \sim 1, \, d \sim 1000 {\rm m}, M_m \sim 450, E_a \sim 80
{\rm kJ mol}^{-1}, C_s \sim 1.0$;
then $\v \= 0.01 - 1000$, $\Lambda=0.1 - 400$, $\ga \= 1$,
$\beta=2.3$ and $\Theta=3.0$.
An initial porosity of $\phi_{0}=0.4$ for pore water and
initial volume fractions of $0.3$ for
montmorillonite, $0.0$ for illite and $0.3$ for quartz are used in
the following computations.
It is worth pointing out that the large variations of $\lambda$ and $\Lambda$
are due to the fact that the permeability $k_0$ varies greatly over a wide
range where the smaller values of $k_0$ correspond to a lower permeability
and larger values correpond to higher peameability. This wide range can thus
similate different types of sediments in different sedimentation environments
(Rieke and Chilingarian, 1974).

\section{Numerical Method}

In order to solve the highly coupled non-linear equations, an implicit
numerical difference method is used. Substituting the expression for
darcy flow  into the other equations, the essential
equations  describing for porosities
$\phi$  and $\phi_m$ become the standard non-linear parabolic form
(Meek and Norbury, 1982)
\begin{equation}
\phi_{t}=F(z,t,\phi) \phi_{zz}+g(z,t,\phi, \phi_{z}).
\end{equation}
The first step gives $\phi^{n+1/2}$ as a solution of the following equation
\[
\frac{2}{\Delta
t}(\phi^{n+1/2}_{i}-\phi^{n}_{i}) \] \[
=(\frac{1}{\Delta
z^2})F(z_{i},t^{n+1/2},\phi^{n}_{i})\delta^{2}_{z} \phi^{n+1/2}_{i} \]
\begin{equation}
+g(z_{i},t^{n+1/2},\phi^{n}_{i},\frac{1}{\Delta
z}\delta_{z} \phi^{n}_{i}),
\end{equation}
where $\delta^{2}_{z}\phi_{i}=(\phi_{i+1}-2\phi_{i}+\phi_{i-1})$ and
$\delta_{z}\phi_{i}=(1/2)(\phi_{i+1}-\phi_{i-1})$. The second stage
gives $\phi^{n+1}_{i}$ as a solution of the following equation
\[ \frac{1}{\Delta
t}(\phi^{n+1}_{i}-\phi^{n}_{i}) \] \[ =(\frac{1}{2(\Delta
z)^2})F(z_{i},t^{n+1/2},\phi^{n+1/2}_{i})\delta^{2}_{z}(\phi^{n+1}_{i}
+\phi^{n}_{i})
\]
\begin{equation}
+g(z_{i},t^{n+1/2},\frac{1}{\Delta
z}\delta_{z} \phi^{n+1/2}_{i}).
\end{equation}
We used a normalized grid parameterized by the
fixed domain variable $Z=z/h(t)$. This will make it possible to compare
in a fixed frame the results at different times
and depths and for different values
of the dimensionless parameters. This transformation maps
the basement of the basin to $Z=0$ and the basin top to $Z=1$.
 Numerical results are
presented and explained in the following section, and are compared
with the real data in the rest of the paper.

\section{Numerical Simulation and Comparison With Real Data}

The main aim of the numerical model is to show how the model equations
behave, to test the validity of the model, and to predict or simulate
real world situations. In principle, a good model should be able to
simulate many realistic features when its physical parameters are
appropriately chosen, but in reality, the accuracy of the simulation
is comprised by a lack of information about the real system.
By changing the different parameters
($\lambda, \Lambda, \beta, \ga, m, n$ and time $t$) as well as the boundary
conditions ($\phi_0$ and heat flux), we can in principle get a best fit
to the real data although such a choice of such parameters is not unique
because different combinations of parameters in the parameter space may
correspond to fit the real data equally well with a same range of
deviations. So the choice of the parameter is still rather arbitrary although
the knowledge of the geology is used in attempt to get a more realistic
grouping of papameters.

The borehole log data (designed as LOG I)  from the DQ-151 borehole
in South China Sea Basin in China was used in the present simulations.
DQ-151 has a depth of 6100 m and its porosity, volume fractions
of smectite and illite, permeability and temperature distribution
were surveyed at nearly 5 meter intervals. The data used below, however, has
been smoothed by averaging over intervals of 250 m, so that
only large scale features are revealed. However the profiles are still
detailed enough to test the model presented in this paper.

The results of the numerical simulations are compared with the data
in Figure 1.  The rescaled height $Z=z/h(t)$
varies from $0$ to $1$ and corresponds to a variation of 6100 m
from the basement to the ocean floor. In simulating
LOG I,  the values of the parameters which best fit the data were
$\v=250$, $\ga=2.4$, $m=8.3$,
$n=1.46$, $\beta=2.4$, $t=7.3$, and $\Theta_c=3.4$. By using the typical
length scale $d=970$ m, $k_0 \=1.22 \times 10^{-16}$ m$^2$, we can
estimate the sedimentation rate at deposition to be  $\dot m_s \=0.77
\times 10^{-11}$ m/s $\=230 $ m Ma$^{-1}$.
The scaling time is about $3.3$ Ma, and thus the dimensional time
is about 24 Ma. For a fixed value of $G=1.0$, the estimated viscosity
is $\xi_0 \approx 5 \times 10^{21}$ N s m$^{-2}$.
We can see that the porosity near the top of the basin
decreases nearly exponentially with depth. It can be approximately
expressed as
\be
\phi=\phi_0 e^{-C_s (h_0-z)},
\ee
where $C_s \=0.93$ is the compaction index. Porosity reduces rapidly
from 0.4 to 0.1. This exponential profile is
consistent with the equilibrium profile given by Fowler and Yang (1998)
found using asymptotic analysis of the nonlinear reaction-diffusion equation
for fast compaction in sedimentary basins.
In the dynamic balance of sedimention and compaction,
the porosity has reached equilibrium in LOG I
in the top 2100 m of the basin.
Fowler and Yang (1998) also predicted that equilibrium even for a compacting
basin may reach a depth of
\be
\Pi=\k{d \ln \v}{m},
\ee
which is about $700 m$ in the present case. The near equilibrium
at the top means that (a) compaction in LOG I is mainly
poroelastic and may reach the equilibrium state at depths much greater
than that predicted by asymptotic analysis, and (b) no overpressuring
occurs in the shallow region above about 2000 m, which is
consistent with the drilling log data.
However, in much deeper regions, the porosity profile
is no longer exponential. A transition occurs at a depth of about 2400 m
below which the porosity becomes nearly uniform. This means
that compaction is in a non-equilibrium state and overpressuring starts to
build up. Compaction gradually becomes
viscous due to the mechanism of pressure solution which operates at
higher pressures and temperatures than those prevailing at depths of less
than 2000 m.

Figure 2 shows the computed temperature profile and the corresponding
data. The values used in this figure are
$\Lambda=150, \, \gamma=0.031$ (degree/m) with the other values being
the same  as in Figure 1.
It is clearly seen that the temperature profile is essentially governed by
conduction  and evolves on a faster time scale than porosity reduction.
This is because $\hat K \pab{\Theta}{z} \sim -1$ when $\Lambda  \gg 1$
and $\hat K \=1$.     Consequently, the temperature distribution is
virtually a linear profile versus depth. Thus we can take
$\Theta=(h-z)$.

Figure 3 compares the computed volume fractions of
different mineral species and with data from the borehole DQ-151.
We can see that the volume fractions of smectite
remains virtually constant from the ocean floor to a depth of about 2100 m,
and suddenly decrease at a depth of about 2400 m, and becomes negligible at
a depth of about 5100 m. Assuming a thermal gradient of nearly 30.7 C/km, the
mineral reaction commenced at nearly $83^{\circ}$C, and it ended
at nearly $160^{\circ}$C. As smectite starts to decrease, the volume fraction
of illite starts to increase and it reaches its maximum at the aforementioned
depth of about 5100 m. The reaction region has a thickness of
 about 2500 m and a temperature variation of nearly $77^{\circ}$C.
The results in this
paper are quite consistent with the general  conclusions drawn
from other data (Abercrombie et al, 1994)
from oceanic and sedimentary basins.

For a linear temperature  distribution with depth (Yang, 2000b)
has analysed the nonlinear equation asymptotically and concluded that
the convection term $\p (\phi_m u^s)/\p z \ll 1$, so that we have
\be
\pab{\phi_m}{t} \= -e^{\beta (h-z-z^*)} \phi_m,
\ee
and thus an approximate solution for the volume fraction
of smectite is
\be
\phi_m =\phi_{m0} \exp[-\k{1}{\beta}
e^{-\beta  (h-z-z^*) } ],
\ee
where $z^*$ is the location of the centre of the reaction region,
and $h$ is the total depth of the basin. This
solution is remarkably accurate in describing the
features within the reaction window. The centre of the reaction window is
\begin{equation}
z^*=\frac{R T^{2}_{0}}{E_{a}} {\rm ln}
\frac{\dot m_{s}}{k^{0}_{r}d }.
\end{equation}
This clearly means that the higher the sedimentation rate $\dot m_s$,
the higher the critical temperature $T^*$ of the mineral reaction,
the deeper the reactive region $z^*$, and vice versa.
A change of 2 orders in sedimentation rate ($\dot m_s$)
will cause a shift of $\Theta_c$ by 2 (equivalently $\sim 60^{0}$ C)
(with other parameters unchanged).
In addition, the thickness of the reaction
region is the order of $(5/\beta) d$.
A typical value of $\beta \approx 2.4 $ gives $2000$ m
(with $d=970$ m), or equivalently a temperature range
of $\sim 63^{0}$C, which is quite close to the range
observed in real data.

\section{Conclusion}

The present model of mechanical compaction, pressure solution,
and the smectite to illite mineral reaction
in hydrocarbon basins uses a rheological relation which incorporates
both poroelastic and viscous effects  and considers mineral
reaction as a one-step dehydration mechanism in a 1-D compacting frame.
 The nondimensional model equations are mainly
controlled by four parameters $\v$, the viscous parameter $\ga$, the
thermal conduction parameter $\Lambda$, and the reaction parameter $\r$.
The highly nonlinear coupled partial differential equations have
been solved using a two stage implicit method
which is quite robust for the present governing equations.

The numerical simulations have shown
that porosity-depth profile is near exponential at shallow depths and then
undergoes a transition to a nearly uniform porosity.
This is because of the
large exponent $m$ in the permeability law $k
=(\phi/\phi_{0})^{m}$, so that even if $\lambda \gg 1$,
the product $\lambda  k$ may  become small at
sufficiently large depths so that the hydroconductivity becomes
very small and thus compaction proceeds extremely slow
(Fowler and Yang, 1998). Meanwhile,
since the bulk viscosity depends on the
porosity in a form of  $\xi=\xi_0 (\phi_0/\phi)^n$ (typically $n \approx 2$),
a decrease to sufficiently low porosity will give an
increase to a very high viscosity.
The transition from the higher porosity to a nearly minimum porosity
is now associated with a sharp change of lower viscosity to a higher
viscosity, and this implies a transition  from poroelastic to
viscous  behavior.
This is because of the
Below  this transition region, the porosity is usually
 uniformly small and viscous compaction essentially dominates.
The sudden switch from poroelastic to viscous compaction is often
associated with a sharp increase in pore pressure and a
low permeability region
where the mineralized seal may be formed. As viscous compaction
proceeds, porosity and  permeability may become so small that
fluid gets trapped below this region, and compaction virtually
stops. Comparison with borehole data has shown very good agreement
between computed porosity profiles and smoothed borehole data.

The temperature evolution and thermal history are mainly controlled by
$\Lambda$, which is the ratio of the time scale of
thermal conduction to the time scale of the sedimentation rate.
Observations suggest that the temperature profile is purely conductive
and evolves on a faster time scale than porosity reduction,
that is to say, $\Lambda \gg 1$ so that $\pab{\Theta}{z} \= -\gamma$
when $\hat K=const$. This is consistent with the linear temperature
profile generated by our model.

The smectite to illite mineral reaction is characterised
by the reaction parameter ${\cal R}$,
which may be defined in terms of a critical temperature $\Theta_{c}$.
This study reveals that  mechanical compaction, which is controlled by
the strata permeability and sedimentation rate, is the most important
geological factor in altering the  porosity evolutions, while
 chemical compaction, controlled by mineral reaction and pressure solution,
causes only small changes in the porosity. The first-order
dehydration model is a good approximation since it
describes the extent of progress of the overall
smectite-to-illite transformation and still reproduces
many essential features of the smectite-to-illite process
if the appropriate reaction rate laws are used based on
the known physics and chemistry from  experimental studies.
Naturally, more work is needed on more realistic formulation
2-D and 3-D compaction and mineral reactions in
sedimentary environments.

{\bf Acknowledgements}: The author would like to thank
the referees, especially
Dr. J A D Connolly and Prof. R A Birchwood,
for their very helpful comments and very instructive suggestions.
I also would like to thank Prof. Andrew C Fowler for his very helpful
direction on mathematical modelling.

\def\Fig#1{\begin{figure} \centerline{\includegraphics[width=3in]{#1}} \caption{}  \end{figure}}

\Fig{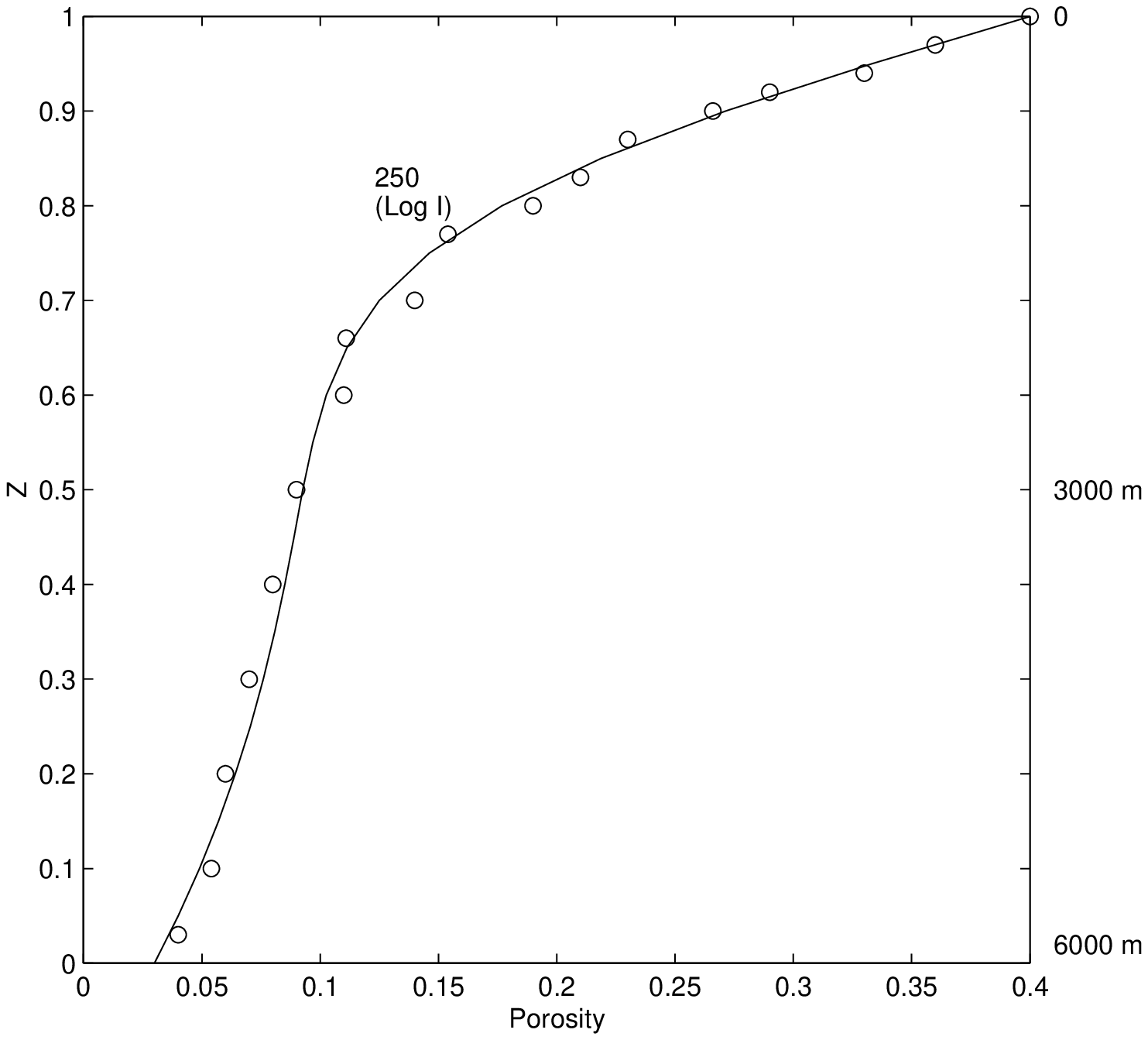}
\Fig{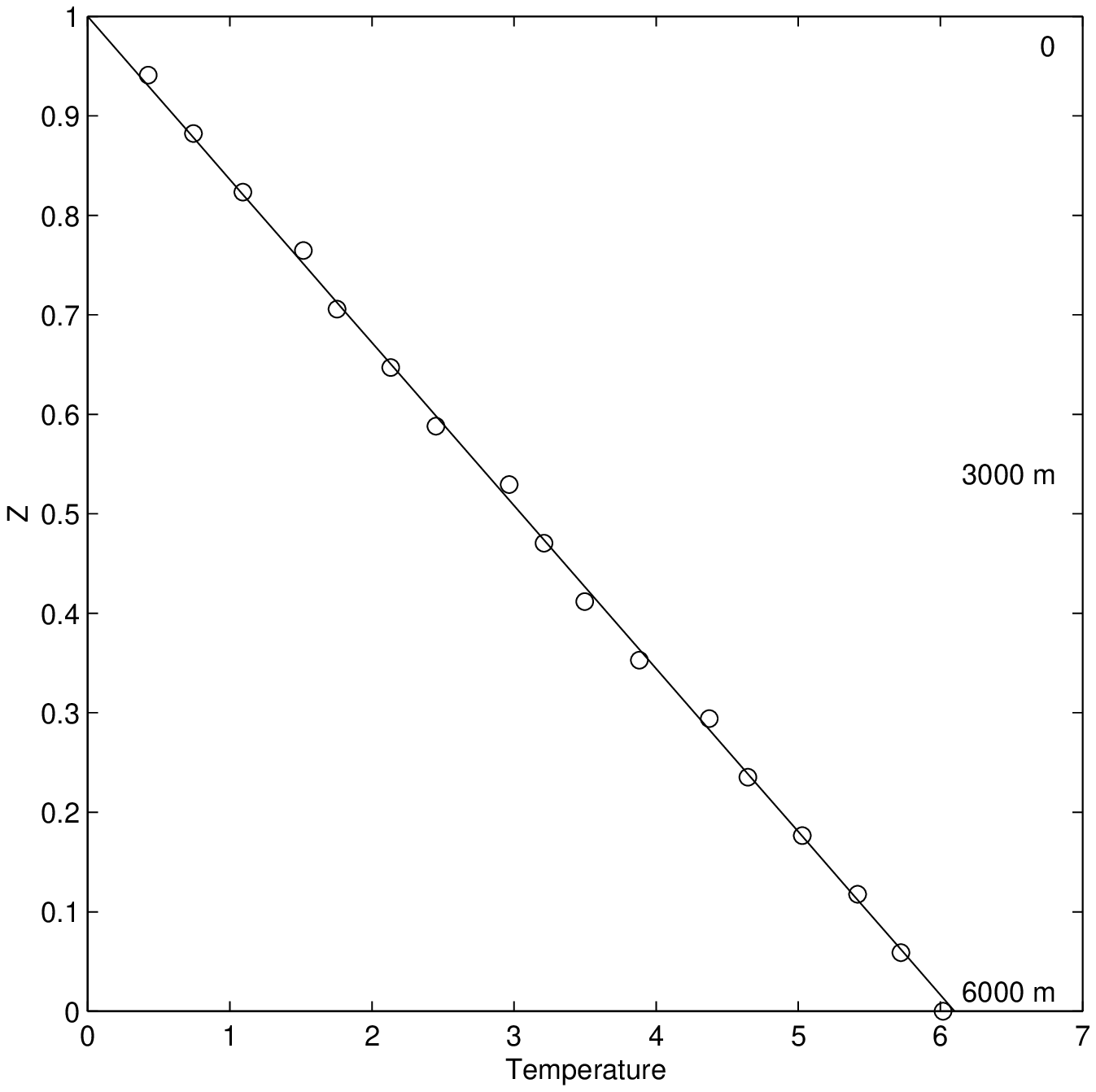}
\Fig{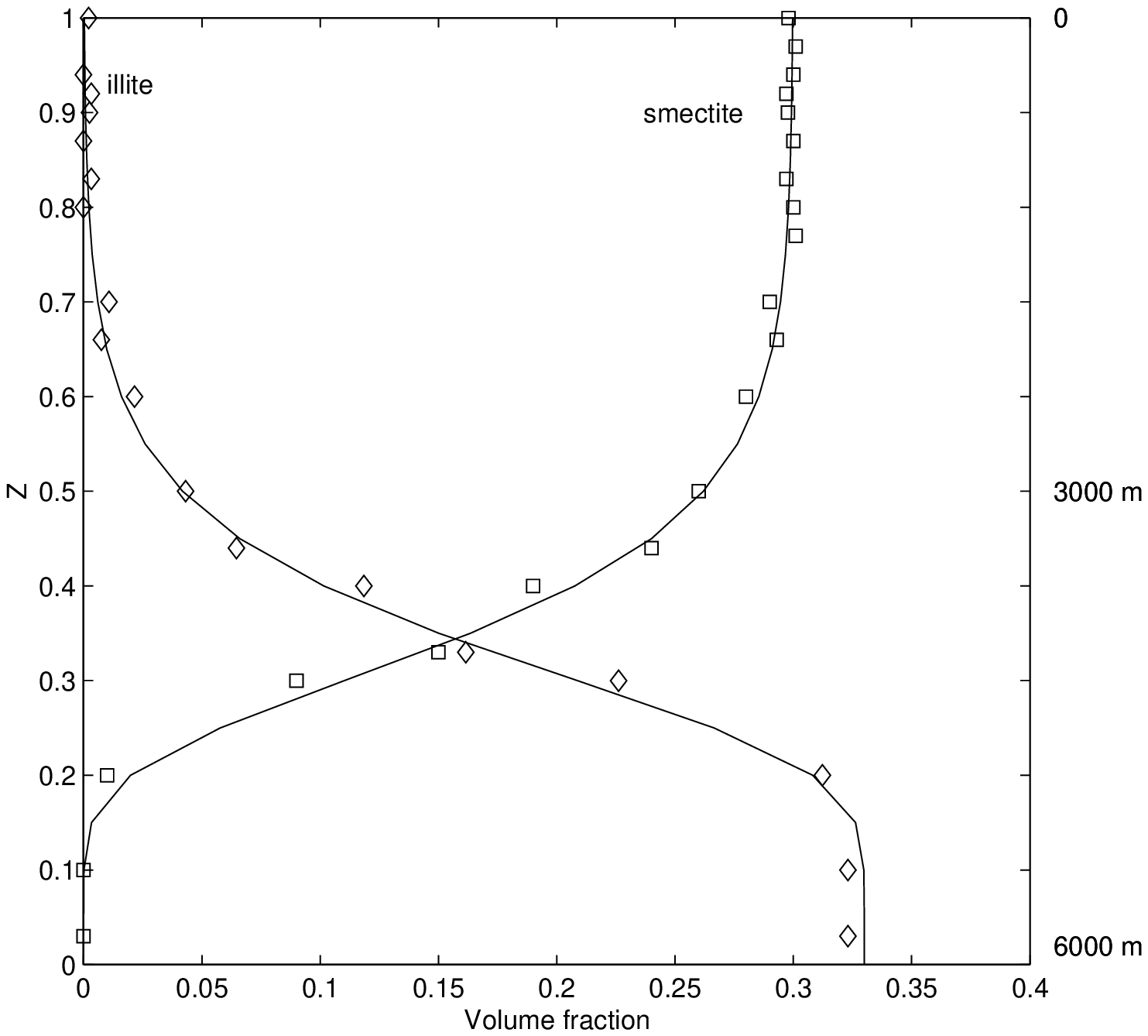}

\subsection*{Figure captions}

\noindent Figure 1. Comparison of computed porosity (solid curve) with
borehole data (depicted as circles) taken from the South China Sea basin. $Z=z / h(t)$ is the scaled height. Compaction in the top region of the basin is essentially at equilibrium (for $\v \=250, \ga=2.4, m=8.3, n=1.46,  t=7.3$).   The porosity profile decreases nearly exponentially.  \\

\noindent Figure 2 Temperature profile for $\Lambda=150$ with
          all other values are the same as in  Fig. 1.     Numerical
          simulation (solid curve) is very consistent
          with the data from Log I.  \\

\noindent Figure 3 Comparison of numerical results (solid curves)
          with data from Log I. The  volume fractions of
          smectite and illite change dramatically in a very thin
          reaction region. The values used in the simulations
          are $\v=250, \, t=7.3,\, \Theta_c=3.4,\, \beta=2.4$.

\section*{REFERENCES}

\begin{description}

\item Abercrombie, H. J, Hutcheon, I. E., Bloch, J. D. \& Caritat, P., 1994,
Silica activity and the smectite-illite reaction, Geology, v.22, p.539-542.

\item  Audet, D.M. \& Fowler, A.C., 1992. A mathematical model for
compaction in sedimentary basins, {\em Geophys. J. Int.}, {\bf 110},
577-590.

\item Angevine, C. L. \& Turcotte, D. L., 1983.  Porosity reduction
by pressure solution: A theoretical model for quartz arenites, {\em
Geol. Soc. Am. Bull.}, {\bf 94}, 1129-1134.

\item Bathurst, R. G. C., 1971. {\em Carbonate sediments and their diagenesis}, Elsevier, Amsterdam.

\item  Birchwood, R. A. \& Turcotte, D. L., 1994. A unified approach to
geopressuring, low-permeability zone formation, and secondary porosity
generation in sedimentary basins, {\em J. Geophys. Res.},
{\bf 99}, 20051-20058.

\item  Bird, R.B., Armstrong, R.C. \& Hassager, O., 1977. Dynamics of
polymeric liquids, Vol.1, John Willy \& Son press.

\item Connolly, J.A.D. and Podladchikov, Y. Y., 2000. Temperature-dependent
viscoelastic compaction and compartmentalization in sedimentary basins,
{\em Tectonophysics}, (in press)

\item Connolly, J. A. D., 1997. Devolatilization-generated fluid pressure
and deformation-propagated fluid flow during regional metamorephism.,
{\em J. Geophys. Res.}, {\bf 102}:18149-18173.

\item Connolly, J.A.D. and Podladchikov, Y. Y., 1998. Compaction-driven fluid
flow in viscoelastic rock. {\em Geodinamica Acta}, {\bf 11}:55-85.

\item  Dunnington, H. V., 1954. Stylolite development post-dates
rock induration, {\em J. Sedimen. Petrol.}, {\bf 24}:27-49

\item   Eberl, D. and Hower, J., 1976, Kinetics of illite formation,
{\em Geol. Soc. Am. Bull.}, {v. 87}, 1326-1330.

\item   Fowler, A.C., 1990. A compaction model for melt transport in the
Earth's asthenosphere. Part I: the basic model, in {\em Magma
Transport and Storage}, ed. Ryan, M.P., John Wiley, pp. 3-14.

\item   Fowler, A. C. and Yang, X. S., 1998. Fast and Slow Compaction
           in Sedimentary Basins, {\it SIAM Jour. Appl. Math.},
           {\bf 59}, 365-385.

\item   Fowler, A. C. and Yang, X. S., 1999.  Pressure Solution and
Viscous Compaction in Sedimentary Basins, {\it J. Geophys. Res.},
B {\bf 104}, 12 989-12 997.

\item  Fowler, A. C., 2000. Compaction and diagenesis,
Proceeding of IMA, (in press)

\item Fuchs, T., 1894. \"Uber die Natur und Entstehung der Stylolithen,
{\em Sitzungsber. Akad. Wiss. Wien. Math-Naturwiss},
{\bf 103}:928-41.

\item   Gibson, R. E., England, G. L. \& Hussey, M. J. L., 1967. The theory of
one-dimensional consolidation of saturated clays, I. finite non-linear
consolidation of thin homogeneous layers, {\em Can. Geotech. J.}, {\bf
17}, 261-273.

\item Heald, M. T., 1955. Stylolites in sandstones,
{\em J. Geol.}, {\bf 63}:101-114

\item   Hedberg, H.D., 1936. Gravitational compaction of clays and shales,
{\em Am. J. Sci.}, {\bf 184}, 241-287.

\item   Lerche, I. 1990. Basin analysis: quantitative methods, Vol. I,
Academic Press, San Diego, California.

\item   McKenzie, D. P., 1984.  The generation and compaction of
partial melts, {\em J. Petrol.}, {\bf 25}, 713-765.

\item   Meek, P.C. \& Norbury, J., 1982. Two-stage, two level finite
difference schemes for no-linear parabolic equations, {\em IMA J. Num.
Anal.}, {\bf 2}, 335-356.

\item  Nield, D. A, 1991. Estimation of the stagnant thermal-conductivity
of the saturated porous media, {\em Int. J. Heat. Mass Trans},
{\bf 34}:1575-1576

\item   Patterson, M. S., 1973. Nonhydrostatic thermodynamics and its
geological applications, {\em Rev. Geophys. Space Phys.}, {\bf 11},
355-389.

\item Revil, A. 1999.  Pervasive pressure-solution transfer: a poro-visco-plastic model, {\em Geophys. Res. Lett.}, {\bf 26}:255-258

item   Rimstidt, J. D. \& Barnes, H. L., 1980. The kinetics of
silica-water reactions,
{\em Geochim. Cosmochim. Acta}, {\bf 44}, 1683-1699.

\item   Rieke, H.H. \& Chilingarian, C.V., 1974. Compaction of argillaceous
sediments, Elsevier, Armsterdam, 474pp.

\item Rutter, E. H., 1983. Pressure solution in nature, theory
and experiment, {\em J. Geol. Soc. London}, {\bf 140}:725-740

\item Schneider, F., Potdevin, J.L., Wolf, S. and Faille, I. 1996.
Mechanical and chemical compaction model for sedimentary basin simulators,
{\em Tectonophysics}, {\bf 263}:307-317

\item   Sharp, J. M., 1976. Momentum and energy balance equations for
compacting sediments, {\em Math. Geol.}, {\bf 8}, 305-332.

\item   Smith, J.E., 1971. The dynamics of shale compaction and evolution in
pore-fluid pressures, {\em Math. Geol.}, {\bf 3}, 239-263.

\item Sorby, H. C., 1863. On the direct correlation of
mechanical and chemical forces, {\em   Proc. R. Soc. London},
{\bf 12}:583-600

\item Stockdale, P. B., 1922. Stylolites: their nature and
origin, {\it Indiana Uni. Stud.}, {\bf 9}:1-97

\item Tada,R. and Siever, R., 1989. Pressure solution during diagenesis,
{\em Ann. Rev. Earth. Planet. Sci.}, {\bf 17}:89-118.

\item   Velde, B. and Vasseur, G., 1992, Estimation of the diagenetic
smectite illite transformation in time-temperature space, { Amer. Mineral.},
{v. 77}, 967-976.

\item   Wangen, M., 1992. Pressure and temperature evolution in sedimentary
basins, {\em Geophys. J. Int.}, {\bf 110}, 601-613.

\item Weyl, P. K., 1959. Pressure solution and the force
of crystallization--a phenomenological theory, {\em J.
Geophys. Res.}, {\bf 64}:2001-2025

\item Yang, X. S., 2000a. Pressure solution in sedimentary basins: effect of temperature gradient, {\em Earth. Planet. Sci. Lett.}, {\bf 176}:233-243

\item Yang, X. S., 2000b. Modeling mineral reactions in compacting sedimentary basins, {\em Geophys. Res. Lett.}, {\bf 27}:1307-1310

\item Yang, X. S., 2000c. Nonlinear viscoelastic compaction in
sedimentary basins, {\em Nonlinear Process Geophys.}, {\bf 7}:1-7

\end{description}

\end{document}